\documentstyle{article}
\begin{document}
\def\ds{\displaystyle}
\def\c{\,\cdot\,}
\def\a{\widehat a}
\def\matr#1#2#3#4{\left(\begin{array}{cc}#1&#2\\#3&#4\end{array}\right)}

\bigskip

\begin{center}
{\Large\bf
Solutions of the functional tetrahedron equation connected with
the local Yang -- Baxter equation for the ferro-electric.
}
\end{center}

\vspace{.5cm}

\begin{center}
S. M. Sergeev\\ \vspace{0.5cm}
Branch Institute for Nuclear Physics, Protvino 142284, Russia.\\
E-mail: sergeev\_ms@mx.ihep.su
\end{center}

\vspace{1cm}

{\small
\noindent
{\bf{Abstract}}:
Local (or modified) Yang -- Baxter equation (LYBE) gives the functional map
from the parameters of the weights in the left hand side to the
parameters of the correspondent weights in the right hand side
of LYBE. Such maps solve the functional tetrahedron equation.
In this paper all the maps associated with LYBE of the ferro-electric type
with single parameter in each weight matrix are classified.
}

\bigskip
\noindent
{\bf{Key words:}} Tetrahedron equation,
$\,2+1\,$ integrability, local Yang -- Baxter equation.
\bigskip

\section{Introduction}

As it is well known, local Yang -- Baxter equations (LYBE-s),
the proper $3D$ generalization of the zero -- curvature condition
\cite{maillet}, are tightly connected with solutions of the so -- called
functional tetrahedron equation.
Usually, LYBE-s are used for the construction of three-dimensional
discrete ``classical'' integrable models
\cite{korepanov-diss,rmk-lybe}. However, in this
paper the field of our interest is the functional tetrahedron equation
itself and a class of its solutions, and
the weights of LYBE we interpret as $L$ -- operators in the
three dimensional sense.

Recall first the general concepts.
Let $L$ be the matrix of the weights, depending on some
independent parameters, say, $\vec x$. The local YBE is
\begin{equation}\label{lybe}
L_{12}(\vec x_a)\;L_{13}(\vec x_b)\;L_{23}(\vec x_c)\;=\;
L_{23}(\vec x'_c)\;L_{13}(\vec x'_b)\;L_{12}(\vec x'_a)\;,
\end{equation}
where $\vec x'_k$ are some functions of $\vec x_k$:
\begin{equation}\label{func}
\vec x'_a = \vec f_1(\vec x_a,\vec x_b,\vec x_c),\;\;\;
\vec x'_b = \vec f_2(\vec x_a,\vec x_b,\vec x_c),\;\;\;
\vec x'_c = \vec f_3(\vec x_a,\vec x_b,\vec x_c).
\end{equation}
It is supposed that:
\begin{itemize}
\item all $L$-s in (\ref{lybe}) have the same functional structure,
and differ only by their vector arguments,
\item being considered as the system of equations with respect to
$\vec x'_k$ (or with respect to
$\vec x_k$), eq. (\ref{lybe}) has the \underline{unique} solution.
\end{itemize}

For given LYBE with the solution (\ref{func}) associate an operator
$R_{a,b,c}$, realizing the map
\begin{equation}
R_{a,b,c}\;:\;\vec x_a,\vec x_b,\vec x_c\rightarrow
\vec x'_a,\vec x'_b,\vec x'_c\;,
\end{equation}
i. e. acting on the space of functions of
$\vec x_a,\vec x_b,\vec x_c$ as follows:
\begin{equation}
R_{a,b,c}\c\phi[\vec x_a,\vec x_b,\vec x_c]\;=\;
\phi[\vec x'_a,\vec x'_b,\vec x'_c]\c R_{a,b,c}\;.
\end{equation}
Then the following formal equation, interpreted as the linear problem
for three dimensional object $R$, holds:
\begin{equation}\label{zam}
L_{12}(\vec x_a)\;L_{13}(\vec x_b)\; L_{23}(\vec x_c)\;\c\;R_{a,b,c}\;=\;
R_{a,b,c}\;\c\;L_{23}(\vec x_c)\;L_{13}(\vec x_b)\; L_{12}(\vec x_a)\;.
\end{equation}
Rather standard manipulations with the quadrilateral formed by six $L$-s
with the arguments $\vec x_1,...,\vec x_6$
allows one to prove from the uniqueness of LYBE the tetrahedron equation
for $R$:
\begin{equation}\label{fte}
R_{1,2,3}\c R_{1,4,5}\c R_{2,4,6}\c R_{3,5,6}\;=\;
R_{3,5,6}\c R_{2,4,6}\c R_{1,4,5}\c R_{1,2,3}\;.
\end{equation}
This is the functional relation, the left and right sides
of it are to be understood acting equivalently
on the space of functions of six vector variables:
\begin{eqnarray}
&\ds R_{123}\c
\Biggl(R_{145}\c
\Bigl( R_{246}\c
\bigl( R_{356}\c
\phi[\vec x_1,...,\vec x_6]\bigr)\Bigr)\Biggr)
\;=&\nonumber\\
&\ds=\;
R_{356}\c
\Biggl(R_{246}\c
\Bigl(R_{145}\c
\bigl(R_{123}\c
\phi[\vec x_1,...,\vec x_6]\bigr)\Bigr)\Biggr)
\;.&
\end{eqnarray}

\section{Local Yang -- Baxter Equation}

Now consider LYBE for simplest two -- state ferro-electric weights:
\begin{equation}
L(a,b,c,d)\;=\;\left(\begin{array}{cccc}
1  & 0  & 0  & 0  \\
0  & a  & b  & 0  \\
0  & c & d & 0  \\
0  & 0  & 0  & z \end{array}\right)\;,
\end{equation}
where the ferro-electric condition $z=bc-ad$.
Being the special case of the free fermionic model,
LYBE for this case can be rewritten in the
equivalent but irreducible form (i.e. in the form where all equations
are independent). This trick is well known, and there is no necessity
do describe it here. Let
\begin{eqnarray}
&\ds X_{12}\;=\;
\left(\begin{array}{ccc}
a_1& b_1&  0\\
c_1& d_1&  0\\
0  & 0  &  1
\end{array}\right)\;,\;\;\;\;
X'_{12}\;=\;
\left(\begin{array}{ccc}
a'_1& b'_1& 0\\
c'_1& d'_1& 0\\
   0&    0& 1
\end{array}\right)\;,&\nonumber\\
&\ds X_{13}\;=\;
\left(\begin{array}{ccc}
a_2& 0& b_2\\
  0& 1&   0\\
c_2& 0& d_2
\end{array}\right)\;,\;\;\;\;
X'_{13}\;=\;
\left(\begin{array}{ccc}
a'_2& 0& b'_2\\
   0& 1&    0\\
c'_2& 0& d'_2
\end{array}\right)\;,&\nonumber\\
&\ds X_{23}\;=\;
\left(\begin{array}{ccc}
 1&   0&   0\\
 0& a_3& b_3\\
 0& c_3& d_3
\end{array}\right)\;,\;\;\;\;
X_{23}\;=\;
\left(\begin{array}{ccc}
1&    0&    0\\
0& a'_3& b'_3\\
0& c'_3& d'_3
\end{array}\right)\;.&
\end{eqnarray}
Thus the irreducible part of LYBE (\ref{lybe}) is
\begin{equation}\label{korepanov}
X_{12}\,\cdot\,X_{13}\,\cdot\,X_{23}\;=\;
X'_{23}\,\cdot\,X'_{13}\,\cdot\,X'_{12}\;.
\end{equation}
Thus there are nine independent relations for twelve ``primed'' variables.
In general the entries of $X_{i,j}$, $a_k,b_k,c_k,d_k$, they are matrices.
Equation (\ref{korepanov}) was investigated by I. G. Korepanov
in this most general form, he proved the irreducibility of this equation
and the existence of an unique (up to some ``gauge'' ambiguities)
solution, and he
pointed out the connection of eq.  (\ref{korepanov}) with
the functional tetrahedron equation first \cite{korepanov-diss}.

Suppose all the elements of $X_k$ and $X'_k$ are numbers in general
position,  so we may extract
form (\ref{korepanov}) the following set of equations:
\begin{eqnarray}
&\ds a_1a_2\;=\;a'_2a'_1\,,\;\;\;\;
{d_1\over z_1}\,{d_2\over z_2}\;=\;
{d'_1\over z'_1}\,{d'_2\over z'_2}\,,&\label{i12}\\
&\ds
d_2d_3\;=\;d'_3d'_2\,,\;\;\;\;
{a_2\over z_2}\,{a_3\over z_3}\;=\;
{a'_2\over z'_2}\,{a'_3\over z'_3}\,&\label{i23}\\
&\ds {z_1\over a_1}\,a_3\;=\;{z'_1\over a'_1}\,a'_3\,,\;\;\;\;
d_1\,{z_3\over d_3}\;=\;d'_1\,{z'_3\over d'_3}\,.&\label{i13}\\
&\ds
z_1z_2z_3\;=\;z'_3z'_2z'_1\,,&\label{i123}
\end{eqnarray}
Now consider the case when $\vec x\equiv x$, this case we call as
one -- parameter functional space. $a,b,c,d$ are
$C$ - number functions of $x$.  In eq. (\ref{korepanov})
there are nine equations, and the form of $X(x)$ is to be chosen so that
there are only three independent entries in (\ref{korepanov}).
This problem is rather nontrivial, the first step of its
solution is considering equations
(\ref{i12},\ref{i23},\ref{i13},\ref{i123}),
this gives several permitted forms of $a(x)$, $d(x)$ and $z(x)$.
The second step is excluding primed variables from rest of
(\ref{korepanov}) and solving the appeared {\em functional} equations
for $b(x)$ and $c(x)$. These manipulations are simple but tedious.

\noindent
{\small
Give just as an example one of the possible reasonings. Suppose
$a(x)$, $d(x)$ and $z(x)$ are nontrivial. Then if the relations
(\ref{i13}) are not equivalent, they would give an expression
for $x'_1$ and $x'_3$ via $x_1$ and $x_3$. If so,
(\ref{i12},\ref{i23},\ref{i123}) must be equivalent,
and this gives either trivial solution $x'_k=x_k$, or contradiction.
Thus both relations in (\ref{i13}) are equivalent, so
one may conclude $z(x)=(1-k)a(x)d(x)$, where $k\neq 1$ is a common
constant. Substituting $c(x)=ka(x)d(x)/b(x)$ into (\ref{korepanov}),
we obtain a functional contradiction quite soon.
Thus the subsequent step is to regard somebody of $a(x)$, $d(x)$ or
$z(x)$ to be a constant, etc.}

\noindent
As the result it appears that there are only six
(up to some equivalence) independent forms of $X$ and so only six
solutions of the functional TE of such kind. We'll numerate them by
the Greece letters.

\section{Solutions of the functional TE}

\subsection{Case $(\alpha)$}

\begin{equation}
\ds X(x)\;=\;\matr{1}{x}{0}{k}\;,
\end{equation}
$k$ being a constant, this gives
\begin{equation}
R_{123}\;:\;x_1,\;x_2,\;x_3\rightarrow x_1,\;kx_2+x_1x_3,\;x_3\;.
\end{equation}
Inverse map:
\begin{equation}
R^{-1}_{123}\;:\;x_1,\;x_2,\;x_3\rightarrow
x_1,\; {x_2-x_1x_3\over k},\;x_3\;.
\end{equation}

\subsection{Case $(\beta)$}

\begin{equation}
\ds X(x)\;=\;\matr{1}{x}{k/x}{0}\;,
\end{equation}
this gives
\begin{equation}
R_{123}\;:\;x_1,\;x_2,\;x_3\rightarrow
{kx_2+x_1x_3\over x_3},\;x_1x_3,\;{kx_2x_3\over kx_2+x_1x_3}\;.
\end{equation}
Inverse map:
\begin{equation}
R^{-1}_{123}\;:\;x_1,\;x_2,\;x_3\rightarrow
{x_1x_2\over x_2+x_1x_3},\; {x_1x_3\over k},\;{x_2+x_1x_3\over x_1}\;.
\end{equation}

\subsection{Case $(\gamma)$}

\begin{equation}
\ds X(x)\;=\;\matr{x}{0}{1-x}{1}\;,
\end{equation}
then
\begin{equation}
R_{123}\;:\;x_1,\;x_2,\;x_3\rightarrow
{x_3-x_2+x_1x_2\over x_3},\;{x_1x_2x_3\over x_3-x_2+x_1x_2},\;x_3\;.
\end{equation}
Inverse map:
\begin{equation}
R^{-1}_{123}\;:\; x_1,\;x_2,\;x_3\rightarrow
{x_1x_2\over x_3-x_1x_3+x_1x_2},\;x_3-x_1x_3+x_1x_2,\;x_3.
\end{equation}

\subsection{Case $(\delta)$}

\begin{equation}
\ds X(x)\;=\;\matr{x}{1}{1-x}{0}\;.
\end{equation}
Then
\begin{equation}
R_{123}\;:\;x_1,\;x_2,\;x_3\rightarrow
{x_1x_2\over x_1+x_3-x_1x_3},\;x_1+x_3-x_1x_3,\;
{(1-x_1)x_2x_3\over x_1+x_3-x_1x_2-x_1x_3}\;.
\end{equation}
Here
\begin{equation}
R^2\;=\;1\;.
\end{equation}
This transformation is connected with the Pentagon equation and
is described in \cite{tenterm}.

\subsection{Case $(\epsilon)$}

\begin{equation}
\ds X(x)\;=\;\matr{x}{1+ix}{1-ix}{x}\;,
\end{equation}
then
\begin{equation}
R_{123}\;:\;x_1,\;x_2,\;x_3\rightarrow
{x_1x_2\over x_1+x_3+x_1x_2x_3},\;
x_1+x_3+x_1x_2x_3,\;
{x_2x_3\over x_1+x_3+x_1x_2x_3}\;,
\end{equation}
again with
\begin{equation}
R^2\;=\;1\;.
\end{equation}
This is the electric network transformation,
considering by R. M.  Kashaev in \cite{rmk-lybe}. Note that he
realized this LYBE in terms of bosonic representation, while
our case is the fermionic one.

\subsection{Case $(\zeta)$}

\begin{equation}
\ds X(x)\;=\;\matr{x}{-s(x)}{s(x)}{x}\;,
\end{equation}
where $s^2(x)=1-x^2$. Thus the argument of $X$ is not simply x, but the
pair $(x,s(x))$.
This case is equivalent to Onsager's star -- triangle and also can
be interpreted
as the decomposition of holonomy group's element with respect to
the Euler basises, and $X$-s are the rotations:
\begin{equation}
\ds X(\phi)\;=\;\matr{\cos\phi}{-\sin\phi}{\sin\phi}{\cos\phi}\;.
\end{equation}
This LYBE was also considered by R. M. Kashaev in \cite{rmk-lybe}.

Formally, in terms  of single $x$, (\ref{korepanov})
gives
\begin{equation}
R_{123}\;:\;x_1,\;x_2,\;x_3\rightarrow
{x_1x_2\over F(x_1,x_2,x_3)},\;
F(x_1,x_2,x_3),\;{x_2x_3\over F(x_1,x_2,x_3)}\;,
\end{equation}
where $F$ can be found from
\begin{equation}
s(F)\;=\;s(x_2)x_1x_3-s(x_1)s(x_3)\;,
\end{equation}
and so this transformation is two -- foiled one (a sign of $F$ is
unessential, there may be two different
signs of $s(x_1)s(x_2)s(x_3)$),
and $R$-s from different foils are inverse.

\subsection{Discussion}

For given map $R_{123}$ there are a lot of equivalent maps and
its descendants. Surely, if $R$ solves the functional TE, then
$R^{-1}$ also solves it. Note that in all cases the inverse maps $R^{-1}$
are connected with $X^t$. Another automorphism is
\begin{equation}
R_{123}\rightarrow R'_{123}\equiv R_{321}\;.
\end{equation}
Also the functional tetrahedron equation admits the gauge freedom
\begin{equation}
R_{123}\rightarrow j_1j_2j_3\circ R_{123}\circ j_1^{-1}j_2^{-1}j_3^{-1}
\;,
\end{equation}
where $j^{-1}$ is the inverse function to $j$:
$j^{-1}(j(x))=j(j^{-1}(x))=x$.

Consider the case when $j(x)=sx$, i.e. one can introduce common
scaling factor for all variables, $x_k\rightarrow sx_k$,
and one can choose it in different
ways (putting it to zero or to something else). Thus from $(\delta)$, as well as
from $(\epsilon)$, one can obtain

\bigskip
\noindent
{\large\bf Case $(\eta)$:}
\bigskip
\begin{equation}
R_{123}\;:\;x_1,\;x_2,\;x_3\;\rightarrow
{x_1x_2\over x_1+x_3},\;x_1+x_3,\;{x_2x_3\over x_1+x_3}\;.
\end{equation}

Also the structure of the tetrahedron equation (\ref{fte})
allows us to impose
an order relation
\begin{equation}
x_1 \;<<\; x_2,x_3\; <<\; x_4,x_5,x_6\;,
\end{equation}
so again $(\delta)$ and $(\epsilon)$ both give

\bigskip
\noindent
{\large\bf Case $(\theta)$}
\bigskip
\begin{equation}
R_{123}\;:\;x_1,\;x_2,\;x_3\rightarrow
x_1{x_2\over x_3},\;x_3,\;x_2\;.
\end{equation}

\vspace{1cm}

\noindent
{\bf Acknowledgments}:
I should like to thank R. M. Kashaev, Yu. G. Stroganov, H. E. Boos,
V. V. Mangazeev, I. G. Korepanov and G. P. Pron'ko
for many fruitful discussions.

\noindent
The work was partially
supported by the grant of the
Russian Foundation for Fundamental research No 95 -- 01 -- 00249.

\end{document}